\documentclass[12pt]{iopart}
\usepackage{graphicx}
\usepackage{dcolumn}
\usepackage{bm}
\usepackage[utf8]{inputenc}
\usepackage[english]{babel}
\usepackage{iopams}
\usepackage{amssymb}
\expandafter\let\csname equation*\endcsname=\relax  
\expandafter\let\csname endequation*\endcsname=\relax 
\usepackage{amsmath}
\usepackage{bbm} 
\usepackage{multirow} 
\usepackage{colortbl} 
\usepackage{booktabs} 

\newcommand{\figurewidth}{0.9\textwidth} 

\begin{document}

\newcommand{\kB}{\ensuremath{k_{\mathrm{B}}}}
\newcommand{\B}{\mathbf{B}}
\newcommand{\Bp}{\ensuremath{B_{\perp}}}
\newcommand{\sigmabf}{\boldsymbol{\sigma}}
\newcommand{\ntd}{\ensuremath{n_{\mathrm{2D}}}}
\newcommand{\EF}{\ensuremath{E_{\mathrm{F}}}}
\newcommand{\kF}{\ensuremath{k_{\mathrm{F}}}}
\newcommand{\aR}{\ensuremath{\alpha_{\mathrm{R}}}}
\newcommand{\bD}{\ensuremath{\beta_{\mathrm{D}}}}
\newcommand{\tg}{\ensuremath{\overline{\overline{g}}^*}}
\newcommand{\mB}{\ensuremath{\mu_{\mathrm{B}}}}

\title[Measurement of Rashba- and Dresselhaus-SOI and $g^*$-factor anisotropy]{Experimental determination of Rashba and Dresselhaus parameters and $g^*$-factor anisotropy via Shubnikov-de Haas oscillations}

\author{F Herzog$^1$, H Hardtdegen$^2$, Th Schäpers$^2$, D Grundler$^1$\footnote{Present address:
École Polytechnique Fédérale de Lausanne, Sciences et Techniques de l'Ingénieur, Institut des Matériaux, Laboratoire des Matériaux Magnétiques Nanostructurés et Magnoniques, 1015 Lausanne, Switzerland} and M~A Wilde$^1$\footnote{Present address: Lehrstuhl für Topologie korrelierter Systeme, Technische Universität München, Physik Department, James-Franck-Strasse 1, D-85748 Garching b. München, Germany} }

\address{$^1$Lehrstuhl für Physik funktionaler Schichtsysteme, Technische Universität München, Physik Department, James-Franck-Strasse 1, D-85748 Garching b. München, Germany}
\ead{mwilde@ph.tum.de}
\address{$^2$Peter Grünberg Institut (PGI-9) and JARA-FIT Jülich-Aachen Research Alliance, Forschungszentrum Jülich, D-52425 Jülich, Germany}

\date{\today}

\begin{abstract}
The spin splitting of conduction band electrons in inversion-asymmetric InGaAs/InP quantum wells is studied by Shubnikov-de Haas measurements combining the analysis of beating patterns and coincidence measurements in doubly tilted magnetic fields. The method allows us to determine the absolute values of the Rashba and linear Dresselhaus spin-orbit interaction coefficients, their relative sign and the full Landé g-tensor. This is achieved by analyzing the \emph{anisotropy} of the beat node positions with respect to both polar and azimuthal angles between the magnetic field direction and the quantum well normal. We show that the spin-orbit interaction is dominated by a large Rashba coefficient together with a linear Dresselhaus coefficient that is 10 $\%$ of the Rashba coefficient. Their relative sign is found to be positive. The g-tensor is found to have a marked out-of-plane anisotropy and a smaller but distinct in-plane anisotropy due to spin-orbit interaction.
\end{abstract}

\pacs{73.21.Fg, 73.43.Qt, 75.70.Tj}

\noindent{\it Keywords}: spin-orbit interaction, Rashba effect, Dresselhaus effect, anisotropic Zeeman interaction, Shubnikov-de-Haas effect, beating patterns, semiconductor spintronics

\submitto{\NJP}
\maketitle

\section{Introduction}
\label{sec:introduction}

Spin related effects in semiconductor two-dimensional electron systems (2DESs) have been subject of intense research in both fundamental physics and research aimed towards novel spintronic devices \cite{Awschalom-2007}. In addition to the anisotropic Zeeman effect relevant for spin manipulation by external magnetic fields \cite{TolozaSandoval2016}, in particular the spin-orbit interaction (SOI) effects in two-dimensional electron systems (2DESs) are of interest. Here, two contributions, i.e., the Rashba (R) \cite{Rashba-1984-2,Manchon2015} effect due to structural inversion asymmetry of the heterostructure and the Dresselhaus (D) \cite{Dresselhaus-1955} effect due to the bulk inversion asymmetry of the crystal play the leading role. The large interest in R-SOI effects in 2DESs is motivated on the one hand by the possibility to manipulate spins by electrical fields \cite{Datta-1990}. On the other hand, both R-SOI and D-SOI can lead to spin decoherence which has to be considered for any device working with spin information \cite{Zutic-2004}. This insight has led to concepts based on the interplay of R-SOI and D-SOI \cite{Schliemann-2003}. In any case, the unambiguous separation and quantification of all these effects in a given electron system is of utmost importance. This is still experimentally challenging. Common methods have addressed in-plane anisotropies of quantities like spin lifetimes that are influenced by SOI. When an external magnetic field is applied, the modelling must include the interplay of SOI and the Zeeman interaction, parameterized by the \emph{anisotropic} Landé tensor $\tg$. The out-of-plane anisotropy of $\tg$ stems from the symmetry reduction in planar heterostructures \cite{Winkler-2003} while the in-plane asymmetry is due to SOI effects in asymmetric heterostructures \cite{Kalevich-1993}. Thus, in asymmetric heterostructures lacking inversion symmetry of the host crystal all of these effects are present simultaneously.

Early studies on SOI effects in 2DESs were performed using magnetotransport measurements. Here, the SOI-induced spin splitting gives rise to distinct beating patterns in the Shubnikov-de Haas (SdH) oscillations \cite{Luo-1988,Luo-1990,Das-1989}. A theoretical model including R-SOI and D-SOI was already formulated in \cite{Das-1990}, although no reference to the resulting in-plane anisotropies was made. Subsequent studies using beating patterns in SdH oscillations focused on samples where a dominant R-SOI was assumed \cite{Nitta-1997,Engels-1997,Grundler-2000,Faniel-2011}. Detailed experimental investigations of SdH beating patterns in tilted magnetic fields and considering both SOI terms are lacking.

Experimental milestones reporting information on  the R-SOI and D-SOI coefficients, denoted by $\aR$ and $\bD$, respectively, include the work by Ganichev \emph{et al} who used the anisotropy of spin photocurrents to determine the ratio $\aR/\bD$ \cite{Ganichev-2004,Giglberger-2007}, without applying a magnetic field, i.e., independent of $\tg$. Meier \emph{et al} performed angle-dependent time-resolved Faraday rotation and extracted the absolute values and relative sign of $\aR$ and $\bD$ but neglected the $\tg$-anisotropy in their analysis \cite{Meier-2007}. Larinov and Golub demonstrated the tunability of $\aR/\bD$ via a gate voltage in angle-dependent time-resolved Kerr rotation \cite{Larionov-2008}. The formation of a persistent spin helix when $\aR=\bD$ was first mapped by Walser \emph{et al} \cite{Walser-2012}.  Recently, Sasaki \emph{et al} performed magnetoconductance measurements on etched nanowires with different in-plane orientations and extracted spin-lifetimes via weak antilocalization (WAL). They demonstrated the tunability of $\aR/\bD$ and the persistent spin helix when $\aR=\bD$ \cite{Sasaki-2014}. To determine $\aR$ and $\bD$ at the same time theoretical works suggest experiments such as electric-dipole spin resonance \cite{Falko-PRB-1992}, magnetoexciton absorption \cite{Olendski-2008} and measurements of the quantum oscillatory magnetization \cite{Wilde-2013} in strong tilted magnetic fields.

Experiments found significant out-of-plane anisotropy of $\tg$ in InGaAs-based systems \cite{Kovalski-1994}, but no substantial out-of-plane anisotropy in AlGaAs-based 2DESs \cite{LeJeune-1997,Malinowski-2000,Pfeffer-2006}. The in-plane anisotropy was quantified in AlGaAs-based systems using spin quantum beat spectroscopy \cite{Oestreich-1996,Eldridge-2011}. An experimental report treating the full $\tg$, as well as R-SOI and D-SOI on the same footing is lacking up to now.

In this paper we report the values of $\aR$, $\bD$, their relative sign and all components of $\tg$ determined on one-and-the-same sample using SdH oscillations detected in magnetotransport on Hall bars in \emph{doubly tilted} magnetic fields $B$. The experiment relies on the anisotropy induced in the node positions of the SdH beatings as a function of the direction of the in-plane component of a magnetic field that is strongly tilted with respect to the 2DES normal. $\aR$, $\bD$ and $\tg$ are determined by fitting model calculations of the node positions to the data. The calculations are based on numerical diagonalization of the single-particle Hamiltonian including R-SOI, and $k$-linear D-SOI terms as well as the anisotropic Zeeman term in an arbitrarily tilted magnetic field.

The paper is organized as follows. In section \ref{sec:theory} we first define the model Hamiltonian. We briefly revisit the so-called coincidence technique \cite{Fang-1968,Nicholas1988} that will later play a role in determining the starting parameters for matching the model to the experiment.  We then outline the numerical calculations and their results with emphasis on the impact of SOI parameters and $\tg$ on the anisotropy of the node positions. In section \ref{sec:methods} we introduce the InP/InGaAs quantum wells (QWs) investigated in this work and describe the experimental setup. We present the experimental results in section \ref{sec:experimental-results} and discuss their implications in section \ref{sec:discussion}. Finally, we draw conclusions in section \ref{sec:conclusion}.

\section{Theory for the analysis of SdH oscillation patterns in doubly tilted fields B}
\label{sec:theory}

We first present the theoretical model that we used to derive the energy states of a 2DES in tilted magnetic fields including $k$-linear Rashba and Dresselhaus SOI as well as anisotropic Zeeman interaction. These energy spectra are key to calculate relevant node positions in magneto-oscillations such as SdH-oscillations as discussed below. We consider an ideal 2DES confined to the $(x,y)$-plane, with $x \parallel [100]$, $y \parallel [010]$ and $z \parallel [001]$. A magnetic field $\B=(B \sin \theta \cos \varphi, B \sin \theta \sin \varphi, B \cos \theta)$ is applied, where $B$ defines the absolute field strength and $\theta$ and $\varphi$ are polar and azimuthal angles, respectively, defining the direction of $\B$. The Hamiltonian $H$ of the problem is now written as follows \cite{Das-1990}:
\begin{align}
\begin{split}
  \label{eq:Hamiltonian}
    H&=H_0+H_{\mathrm{Z}}+H_{\mathrm{R}}+H_{\mathrm{D}}\\
    H_0&=\frac{\boldsymbol{\pi}^2}{2 m^*} \\
    H_{\mathrm{Z}}&=\frac{1}{2} \mB \ \sigmabf \cdot \tg \cdot \B \\
    H_{\mathrm{R}}&=\frac{\aR}{\hbar}(\sigma_x \pi_y - \sigma_y \pi_x) \\
    H_{\mathrm{D}}&= \frac{\bD}{\hbar}(\sigma_x \pi_x - \sigma_y \pi_y)
    \end{split}
\end{align}
Here, $H_0$ is the orbital part with the kinetic momentum $\boldsymbol{\pi}=\hat{\mathbf{p}}+e\mathbf{A}$, $\hat{\mathbf{p}}=(\hat{p}_x,\hat{p_y})$, the vector potential $\mathbf{A}$ defining the magnetic field $\B= \nabla \times \mathbf{A}$ and the effective mass $m^*$. $H_{\mathrm{Z}}$ defines the Zeeman interaction expressed with the Bohr magneton $\mB$, the spin operator $\sigmabf=(\sigma_x,\sigma_y,\sigma_z)$ and the effective Landé tensor $\tg$. The terms $H_{\mathrm{R}}$ and $H_{\mathrm{D}}$ denote $k$-linear R- and D-SOI terms with SOI parameters $\aR$ and $\bD$, respectively. Note that we neglected the influence of the in-plane field component $B_{\parallel}=B \sin \theta$ on the orbital movement in $z$-direction. This is justified as long as the energy separation of the 2D subbands in the QW is much larger than $\hbar e B_{\parallel}/ m^*$ \cite{Wilde-2009}, which was the case in our experiments presented in section~\ref{sec:results} ($\hbar=h/(2\pi)$ is the reduced Planck constant).

Following Ref.~\cite{Das-1990}, we calculate the matrix elements $H_{ms_z,ns_z'}=\langle m, s_z| H | n, s_z' \rangle$ of $H$ in terms of the eigenstates $|n,s_z \rangle$ of $H_0$ ($n=0,1,2,.. , s_z = \pm$) and diagonalize $H$ numerically by truncating at a finite $n_{\text{max}}$, yielding the eigenstates $|n'\rangle$ with energies $E_{n'}$ at a particular field $\B$. In the following, we will use the perpendicular field component $\Bp=B \cos \theta$ together with values of $\theta$ and $\varphi$ to discuss $E_{n'}(\Bp)$.

\begin{figure}
  \centering
  \includegraphics[width=\figurewidth]{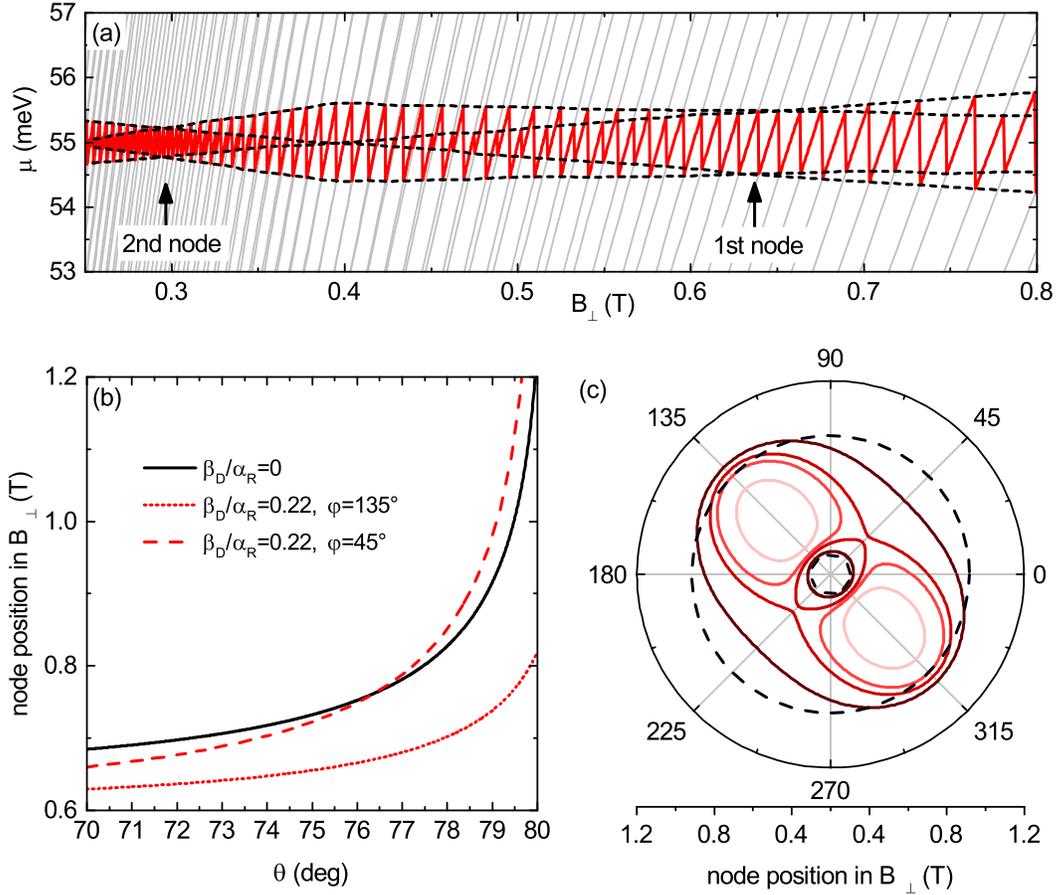}
  \caption{(Color online) (a) Calculated energy levels $E_{n'}(\Bp)$ (grey lines) and chemical potential $\mu (\Bp)$ (solid red line) at constant $\ntd$ including R-SOI for perpendicular field orientation ($\theta=0^{\circ}$). Simulation parameters were $m^*=0.037 m_0$, $g^*=-4.45$ (isotropic), $\aR=4.5 \times 10^{-12} \ \mathrm{eVm}$, $\bD=0$ and $\ntd=8.5 \times 10^{11} \ \mathrm{cm^{-2}}$. The dashed lines are the two envelopes of the quantum oscillations in $\mu (\Bp)$ for the sets of even and odd integer $\nu$, respectively. Beat node positions for the first and second node are indicated by arrows. (b) Position of the first beat node in $\Bp$ as a function of the polar angle $\theta$ for $\bD=0$ (solid black line) and for $\bD/\aR=0.22$ at $\varphi=45^{\circ}$ (dotted red line) and at $\varphi=135^{\circ}$ (dashed red line). $\aR$ was fixed as in (a). (c) $\varphi$-polar plots of first and second node positions at fixed $\theta=78^{\circ}$ and $\aR$ as in (a) for $\bD/\aR=\{0.22,0.34,0.44,0.55 \}$ (from dark to light). The dashed line represents the isotropic case of $\bD=0$. }
  \label{fig:theory_isotrop}
\end{figure}

We first concentrate on the case of isotropic Zeeman interaction, i.e., $\tg=g^*\mathbbm{1}$ ($g^*$ is the scalar Landé factor and $\mathbbm{1}$ is the identity matrix). Figure \ref{fig:theory_isotrop} (a) displays the calculated $E_{n'}(\Bp)$ for perpendicular field orientation, i.e., $\theta=0$, using parameters of the investigated InP/InGaAs material system including R-SOI, but first neglecting D-SOI.
 The chemical potential $\mu$ at constant sheet electron density $\ntd=8.5 \times 10^{11} \, \mathrm{cm^{-2}}$ and zero temperature is indicated in red. $\mu$ exhibits field-dependent quantum oscillations where the sharp jumps correspond to integer values of the filling factor $\nu=\ntd/(e \Bp/h)$. The presence of R-SOI leads to beatings in the quantum oscillations in such a way that the peak-to-peak amplitudes of the jumps at even and odd integer $\nu$ interchange their signal strength (dashed lines in figure~\ref{fig:theory_isotrop} (a)). Neglecting level broadening and a finite temperature for the moment these amplitudes directly correspond to the energy gap between subsequent spin-split Landau levels (LLs) $E_{n'}$ \cite{Wilde-2008,Wilde-2014,MacDonald-1986}.

We now define a beat node position at field values where $(E_{n'}-E_{n'-1})=(E_{n'+1}-E_{n'})$, i.e., where energy gaps for subsequent $\nu$ are equal in size. In the following analysis, we concentrate on the first two beat node positions, as indicated in figure \ref{fig:theory_isotrop} (a). When the magnetic field is increasingly tilted away from the surface normal, the node positions move to larger $\Bp$ once the ratio of Zeeman energy and SOI strength becomes considerable. This is displayed in figure \ref{fig:theory_isotrop} (b), where the position of the first node is plotted as a function of $\theta$. The solid black line represents the case where $\aR>0$ and $\bD=0$ and revisits the results already presented by Das \emph{et. al.} \cite{Datta-1990}. As $\theta$ is increased, the node moves to larger values in $\Bp$. At high field values, the Zeeman interaction becomes dominant over SOI and the curve asymptotically approaches the angle of half coincidence $\theta_c^{1/2}$, which is defined as \cite{Nicholas1988}
\begin{equation}
  \label{eq:theta_c}
   \cos \theta_c^{1/2}=\frac{2 g^* \mB m^*}{e \hbar} \ .
\end{equation}
Note that the result shown as the solid line in figure~\ref{fig:theory_isotrop}~(b) is valid for all $\varphi$, i.e., the node position is isotropic for all azimuthal angles. Now, if we additionally assume a non-zero D-SOI contribution with $\bD \neq 0$, the node positions are no longer isotropic, but depend on the azimuthal orientation $\varphi$ of $\B$. This is illustrated by the dashed and dotted lines in figure \ref{fig:theory_isotrop} (b), which represent the extremal cases of $\varphi=45^{\circ}$ with $\B_{\parallel} \parallel [110]$ and $\varphi=135^{\circ}$ ($\B_{\parallel} \parallel [1\bar{1}0]$). The anisotropy becomes most significant when $\theta$ approaches $\theta_c^{1/2}$. To illustrate this anisotropy, we plot the node positions at fixed $\theta$ in a $\varphi$-polar plot in figure \ref{fig:theory_isotrop} (c) for different $\bD$ at fixed $\aR$. The $\varphi$-anisotropies of the first and the second beat node have a different relative sign (not shown), i.e., the maxima in $\Bp (\varphi)$ for the first node correspond to the minima for the second node. The patterns are rotated by $90^{\circ}$. As the ratio $\bD/\aR$ increases, the anisotropy for both nodes gets larger and both nodes move one toward another until they meet at $\varphi=45^{\circ}$. When $\bD/\aR$ is further increased the nodes vanish for larger and larger ranges of $\varphi$. If $\bD/\aR$ approaches one, our model shows that the nodes vanish for the whole $\varphi$-range. This behavior agrees with previous works that predict the absence of beating patterns in magneto-oscillations if $\aR=\bD$ \cite{Tarasenko-2002}.

A reversal of the relative sign of $\aR$ and $\bD$ rotates the anisotropy patterns in $\varphi$ by $90^{\circ}$. We note that the combined $\varphi$-evolution of both node positions as displayed in figure \ref{fig:theory_isotrop} (c) is \emph{unique} for a particular set $(\aR,\bD)$, i.e., there is no second set that leads to the same curves. This means that a measurement of $\varphi$-dependent node positions gives access to both absolute values \emph{and} relative sign of $\aR$ and $\bD$. Solely the absolute signs of $\aR$ and $\bD$ remain ambiguous, meaning that the cases $(\aR,\bD)$ and $(-\aR,-\bD)$ cannot be distinguished.

\begin{figure}
   \centering
    \includegraphics[width=\figurewidth]{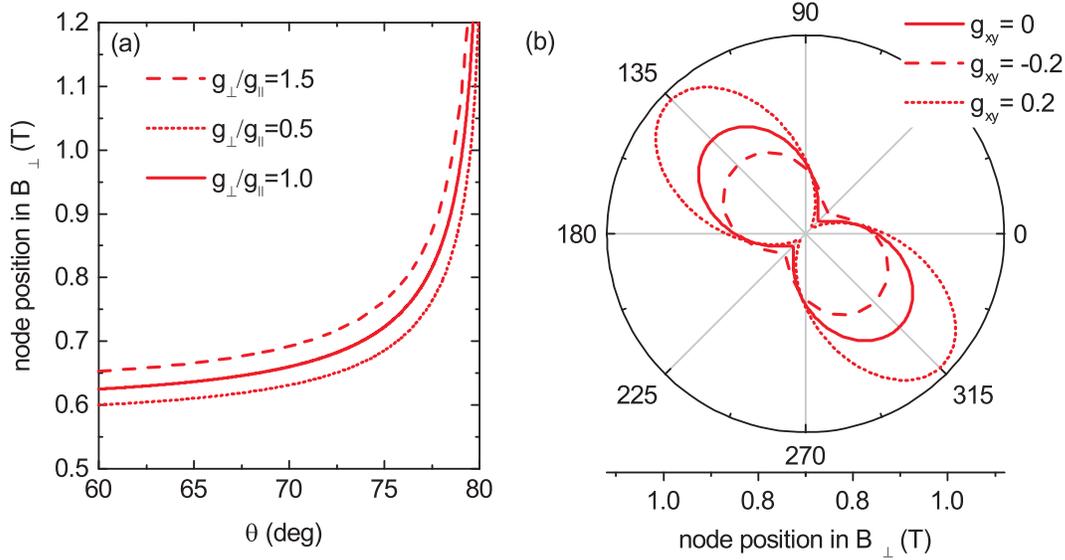}
   \caption{Calculated position in $B_{\perp}$ of the first beat node considering different strengths of the $\tg$-anisotropy. (a) $\theta$-dependence for different $q=g_{\perp}/g_{\parallel}$. Simulation parameters were $m^*=0.037 m_0$, $\aR=4.5 \times 10^{-12} \ \mathrm{eVm}$, $\bD=0.99 \times 10^{-12} \ \mathrm{eVm}$, $\ntd=8.5 \times 10^{11} \ \mathrm{cm^{-2}}$ and $\theta_c^{1/2}=80.136^{\circ}$. (b) $\varphi$-dependence for different values of $g_{xy}$. Parameters were set as in (a) and $q=1$ (corresponding to $g_{\perp}=g_{\parallel}-4.63$) was fixed.}
   \label{fig:theor-g-anisotr}
 \end{figure}

Up to now we have neglected a possible anisotropy of the $\tg$-tensor. This anisotropy could influence the results above, since the node positions result from the interplay of R-SOI, D-SOI $\emph{and}$ the Zeeman energy. Besides the out-of-plane anisotropy of $\tg$, which naturally occurs in narrow (001)-oriented QWs \cite{Ivchenko-1992,Pfeffer-2006}, the combined effect of R-SOI and D-SOI provokes an in-plane anisotropy of $\tg$ \cite{Kalevich-1993,Nefyodov-2011}. In order to provide a realistic description of the system we take the full $\tg$-tensor into account in the numerical modeling:
\begin{equation}
  \label{eq:g-tensor}
  \tg=\begin{pmatrix} g_{\parallel} & g_{xy} & 0 \\ g_{xy} & g_{\parallel}& 0 \\ 0 & 0 & g_{\perp} \end{pmatrix} \, .
\end{equation}
Here, $g_{\parallel}$ and $g_{\perp}$ denote the in- and out-of-plane components of $\tg$, while the off-diagonal term $g_{xy}$ due to SOI parameterizes the in-plane anisotropy. Figure \ref{fig:theor-g-anisotr} (a) shows calculated field positions of the first beat node as a function of $\theta$ for different $q=g_{\perp}/g_{\parallel}$ ($g_{xy}=0$) and fixed $\theta_c^{1/2}$.
 The model reveals that the node positions at a particular $\theta$ are significantly shifted as a function of $q$, implying that the consideration of out-of-plane anisotropy of $\tg$ is relevant for a correct extraction of the SOI-constants. Furthermore, in figure \ref{fig:theor-g-anisotr} (b), the effect of a non-zero $g_{xy}$ on the $\varphi$-anisotropy of the node positions is demonstrated. Qualitatively speaking, $g_{xy}$ provokes a $\varphi$-anisotropy of $\theta_c^{1/2}$, shifts the asymptotes of the curves shown in (a) and thus leads to an enhancement or a suppression of the anisotropy in the node positions due to SOI.

In order to fit the model to the experimental data it is useful to try and find reasonable starting values and reduce the number of free parameters in the fitting routine. This can be achieved by, e.g., measuring the coincidence angles $\theta_c^{1/2}$ separately at higher fields where the influence of SOI is smaller or even negligible. If we neglect the SOI terms $H_{\mathrm{R}}$, $H_{\mathrm{D}}$ in (\ref{eq:Hamiltonian}) (which is justified at large $\Bp$) the Hamiltonian can be diagonalized analytically, yielding a direction-dependent scalar $g^*$-factor of the form
\begin{equation}
  \label{eq:g-factor-polar}
g^*(\theta,\varphi)=\sqrt{g_{\perp}^2\cos^2 \theta+(g_{\parallel}^2+g_{xy}^2) \sin^2 \theta + 2g_{\parallel}g_{xy}\sin^2 \theta \sin 2\varphi}  \ .
\end{equation}
Note that $\tg$ is diagonal in the basis where $x$ and $y$ axes are parallel to $[110]$ and $[1 \bar{1} 0]$, respectively. Given the fact that $g_{xy}$ is expected to be small, i.e., $|g_{xy}|<< (|g_{\parallel}|,|g_{\perp}|)$ \cite{Kovalski-1994,Oestreich-1996}, the out-of-plane and in-plane dependencies can be simplified to
\begin{equation}
  \label{eq:g-polar-simple}
   g^*(\theta)=\sqrt{g_{\perp}\cos^2{\theta}+g_{\parallel} \sin^2 \theta} \quad \mathrm{and} \quad g^*(\theta=90^{\circ},\varphi)=g_{\parallel} + g_{xy} \sin{2\varphi} \ .
\end{equation}
We stress that this analytical result is not expected to hold at small $\Bp$ in the presence of R-SOI and D-SOI terms, and numerical calculations of $E_{n'}(\Bp)$ are required in this case. However, equation (\ref{eq:g-factor-polar}) together with (\ref{eq:theta_c}) provides a tool to analytically calculate the coincidence angle $\theta_c^{1/2}$, which is approximately independent of R-SOI and D-SOI terms at large $\Bp$.

In conclusion, we find that direct access on SOI-constants via a measurement of $\varphi$-anisotropies in the node positions is only possible by considering the full anisotropy of $\tg$. The model suggests that this can be achieved by accessing these anisotropies at different values of $\theta$ and by addressing the respective coincidence angles $\theta_c^{1/2}$ separately at the highest achievable field. Such measurements are suitable to extract both R- and D-SOI constants \emph{and} the full anisotropy of the Landé tensor by careful fitting to the model calculations.

\section{Materials and methods}
\label{sec:methods}
\begin{figure}
   \centering
    \includegraphics[width=0.7\textwidth]{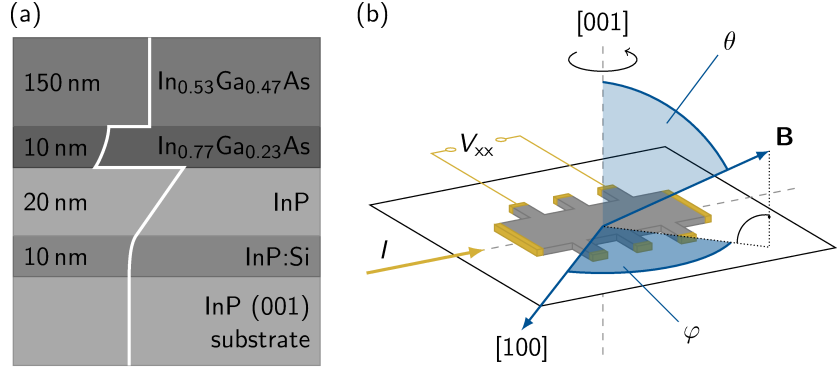}
   \caption{(a) Layer sequence of the InP/InGaAs heterostructures. The energy of the conduction band edge is indicated schematically by the solid line. (b) Sketch of the Hall-bar geometry including the crystallographic orientation. The direction of the applied magnetic field $\B$ with polar and azimuthal angles $\theta$ and $\varphi$ is indicated. }
   \label{fig:exp-setup}
 \end{figure}

The 2DESs investigated here were formed in modulation-doped 10 nm-thick asymmetric In$_{0.77}$Ga$_{0.23}$As quantum wells (QWs) sandwiched between InP and InGaAs barrier layers grown by metal-organic vapor phase epitaxy. The QWs were strained and optimized for, both, high electron mobility and dominant R-SOI \cite{Hardtdegen-1993}. A sketch of the layer sequence with a schematic band diagram of the conduction band energy is displayed in figure~\ref{fig:exp-setup} (a).

Electrical characterization was performed at $T=0.3$~K on standard $(700 \times 200) \, \mathrm{\mu m^2}$ Hall-bar mesas with AuGe contacts [figure \ref{fig:exp-setup} (b)]. These Hall bars were oriented along the $[1 \bar{1} 0]$-direction in the 2DES plane. Longitudinal and transverse resistance, $R_{xx}$ and $R_{xy}$, were recorded simultaneously using standard lock-in techniques. The magnetic field was applied with tilt angles $\theta$ and $\varphi$ defined in section \ref{sec:theory}. Prior to all measurements presented here, the samples were illuminated \emph{in situ} with a blue light emitting diode for 10 s in order to maximize $\ntd$ via the persistent photoeffect. Two different samples, \#4069-6 and \#4069-7, were investigated, leading to consistent results.

Coincidence angle measurements were performed in a 15~T vertical axis superconducting magnet using a mechanical rotator sample stage with the rotation axis normal to the magnetic field direction. In these experiments, $\theta$ was adjusted \emph{in situ} for two different fixed in-plane orientations $[1 \bar{1} 0]$ ($\varphi=135^{\circ}$) and $[1 1 0]$ ($\varphi=45^{\circ}$) addressed in subsequent cool-down cycles. High-field $R_{xx}$-oscillations were recorded for different $\theta$ in the vicinity of the coincidence angle ($\approx 80^{\circ}$). The angle $\theta$ between magnetic field and Hall bar normal was calibrated using the low-field Hall resistance $R_{xy}$.

The anisotropy of the beat node positions was measured using a 2-axis vector magnet with 4.5~T rotatable field magnitude. This allowed us to adjust $\theta$ by electronic means and achieve a very high angular accuracy. $\varphi$ was adjusted by mechanically rotating the sample \emph{in situ} about its vertical axis [see figure \ref{fig:exp-setup} (b)]. Quantum oscillations of $R_{xx}$ were recorded for three different $\theta=\{76.5^{\circ}, 77.5^{\circ}, 78.5^{\circ}\}$ and different values of $\varphi$. The absolute accuracy of $\theta$ at different values of $\varphi$ was $0.07^{\circ}$ as measured using the Hall resistance.

\section{Results}
\label{sec:results}

\subsection{Experimental results}
\label{sec:experimental-results}

\begin{figure}
  \centering
  \includegraphics[width=\figurewidth]{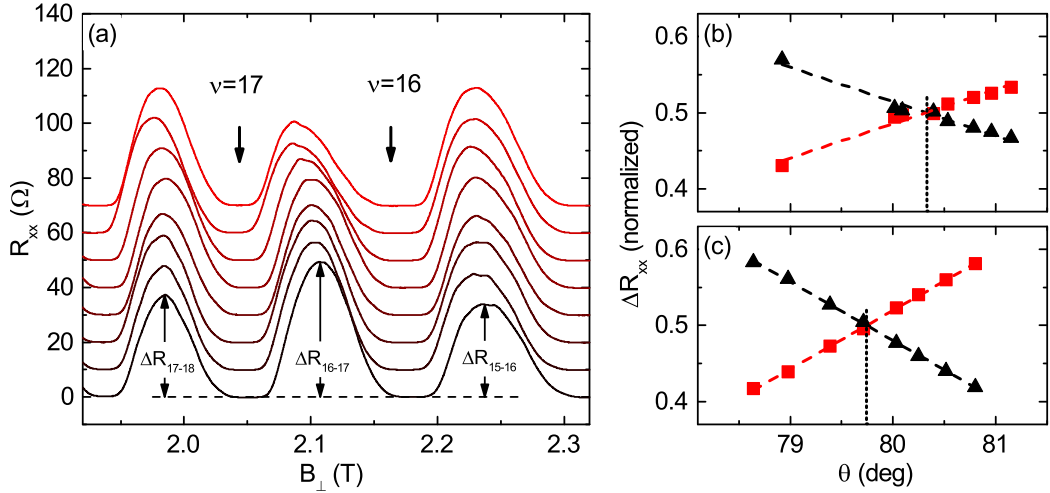}
  \caption{(Color online) Coincidence measurements. (a) $R_{xx}$-traces in the vicinity of $\Bp=2.1 \, \mathrm{T}$ at $\varphi=135^{\circ}$ and different $78.6^{\circ}<\theta <80.8^{\circ}$ (from light to dark). Curves are offset for clarity. (b) and (c) Normalized amplitudes of $\Delta R_{16-17}$ (red squares) and $1/2 (\Delta R_{15-16}+ \Delta R_{17-18})$ (black triangles) recorded at $\varphi=45^{\circ}$ and $\varphi=135^{\circ}$, respectively. The dashed lines are linear fits to the data; the coincidence angle $\theta_c^{1/2}$ was determined by the crossing points of these fits (indicated by vertical dotted lines).}
  \label{fig:exp-rxx-coincidence}
\end{figure}

We first address the coincidence measurements that we used to gain information on the Landé tensor and supply a starting point for fitting the full numerical model to the experimental data. $R_{xx}$-traces for different $\theta$ around $\theta_c^{1/2}$ and fixed $\varphi=135^{\circ}$ ($B_{\parallel} \parallel [1 \bar{1} 0]$) are shown exemplarily in Figure \ref{fig:exp-rxx-coincidence} (a).
 The maximum perpendicular field component in this angular range was $\Bp < 2.35 \, \mathrm{T}$, corresponding to filling factors $\nu>15$. In order to extract $\theta_c^{1/2}$, we plotted the normalized amplitudes $(\Delta R_{15-16}+\Delta R_{17-18})/2$ and $\Delta R_{16-17}$ as a function of $\theta$ in figure \ref{fig:exp-rxx-coincidence} (b) and (c) and defined $\theta_c^{1/2}$ as the crossing point of linear fits to the data. Using this approach it was possible to extract $\theta_c^{1/2}$ with an uncertainty of $\Delta \theta \approx 0.1^{\circ}$.

For the following analysis we first neglect the R-SOI and D-SOI terms as discussed above such that the picture of an angle-dependent scalar $g^*(\theta,\varphi)$ is applicable. Comparing the results for $\varphi=45^{\circ}$ and $\varphi=135^{\circ}$ we find an anisotropy of $\theta_c^{1/2}$, corresponding to different values of $g^*$ (see table \ref{tab:coincidence_angles}).
\Table{ \label{tab:coincidence_angles} Experimental coincidence angles $\theta_c^{1/2}$ and calculated scalar $g^*$
    for sample \#4069-7.}
\noindent  \begin{tabular}{clcc} \br
    $\varphi$ {\footnotesize (deg)} & $\theta_c^{1/2}$ {\footnotesize (deg)} & $g^*$ \\\mr
    $45$ & $80.33 \pm 0.1$ & $-4.54 \pm 0.05$ \\
    $135$ & $79.74 \pm 0.1$ & $-4.81 \pm 0.05$ \\\br
  \end{tabular}
\endTable In this first analysis neglecting R-SOI and D-SOI terms this anisotropy is entirely attributable to a non-zero $g_{xy}$. Note that with this ansatz we will extract an \emph{upper} bound for $g_{xy}$, since the SOI terms that we neglected here produce an anisotropy with the same symmetry. This will be discussed in more detail below. With this approach, two of three independent components of $\tg$ can be determined. The third is lacking because the $\theta$ dependence is not addressed separately in a coincidence measurement.
Since the deviations from in-plane isotropy are small we can use the following approximations: In eq. (\ref{eq:g-factor-polar}), we set $\theta \approx 80^{\circ}$ and substitute $g_{80} = g_{\parallel} \sqrt{q^2 \cos^2 80^{\circ} + \sin^2 80^{\circ}}$. Here, the out-of-plane anisotropy of $g^*$ is represented by the unknown parameter $q=g_{\perp} /g_{\parallel}$. Since $q$ is on the order of one, we further approximate in eq. (\ref{eq:g-polar-simple}) $g_{\parallel} \approx g_{80}$ yielding
\begin{equation}
    \label{eq:g80phi}
    g^*(\theta=80^{\circ},\varphi)\approx g_{80}+g_{xy} \sin 2 \varphi \ .
\end{equation}
We extract $g_{80}= -4.68 \pm 0.04$ and the upper bound $g_{xy}=0.14 \pm 0.04$, which we will use as starting points for the modeling below.

\begin{figure}
  \centering
  \includegraphics[width=\figurewidth]{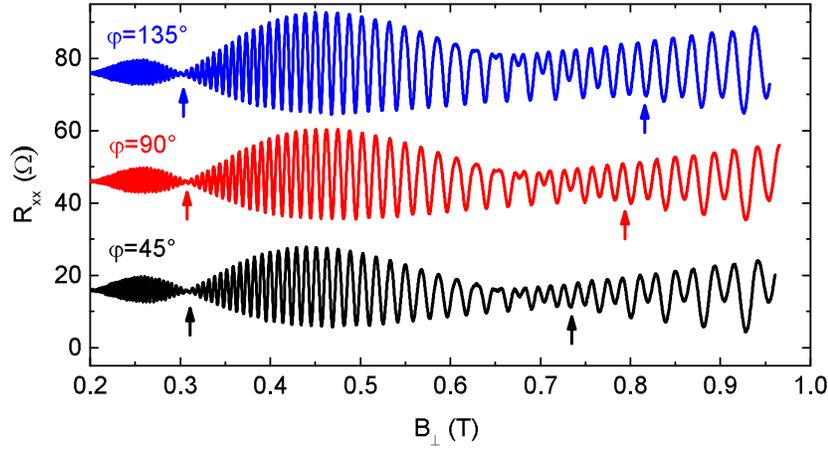}
  \caption{Experimental $R_{xx}$-traces as a function of $\Bp$ of sample \#4069-7 recorded at $\theta=77.5^{\circ}$ at different in-plane orientations $\varphi$. Curves are offset for clarity. The positions of the first two beat nodes are indicated by arrows.}
  \label{fig:exp-rxx-polar}
\end{figure}

\begin{figure}
  \centering
  \includegraphics[width=\figurewidth]{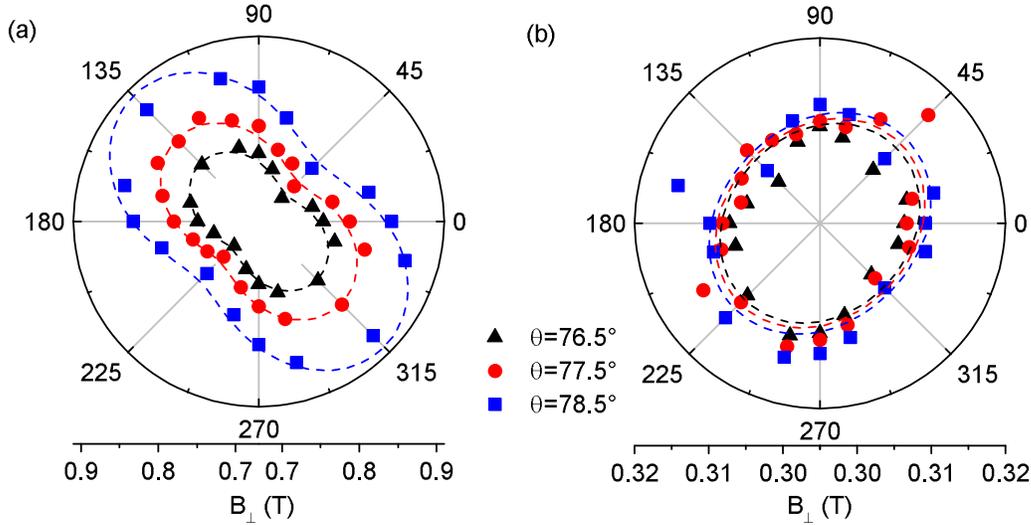}
  \caption{Polar plots in $\varphi$ of the experimental node positions (symbols) of (a) first and (b) second node at three different values of $\theta$. The dashed lines represent the fit of the theoretical model to the data, taking into account R- and D-SOI as well as the full anisotropic $\tg$-tensor.}
  \label{fig:nodepos-polar}
\end{figure}

We now turn to the anisotropy of the measured beat node positions. We show $R_{xx}$-traces of sample \#4069-7 for constant $\theta=77.5^{\circ}$ and three different $\varphi= \{45^{\circ},90^{\circ}, 135^{\circ}\}$ in Figure \ref{fig:exp-rxx-polar}.
 The beat nodes as defined in section \ref{sec:theory} are indicated by arrows. The nodes shift as a function of $\varphi$. The node positions were extracted from the crossing points of the amplitude evolution for even and odd $\nu$. This procedure resulted in an uncertainty of 10 mT for the first node and 3 mT for the second node, while the uncertainty in $\theta$ for the different $\varphi$ was negligible. The experimental node positions for sample \#4069-7 are summarized in $\varphi$-polar-plots in Figure \ref{fig:nodepos-polar} (a) and (b).
 We observe a clear $\varphi$-anisotropy of the first node which becomes significantly stronger with increasing $\theta$. The anisotropy of the second node is significant but less pronounced and exhibits an anisotropy pattern that is rotated by $90^{\circ}$ with respect to the first node, consistent with our theoretical model.

\subsection{Comparison with theory}
\label{sec:comp-with-theory}

In the following we will use the theoretical model introduced in section \ref{sec:theory} to extract all relevant parameters $\aR$, $\bD$, $g_{\perp}$, $g_{\parallel}$ and $g_{xy}$ in a self-consistent way.
For this, we first perform a least-squares fit of the full theoretical model (including $\aR$, $\bD$ and $\tg$) to the data of \emph{all} experimental node positions, i.e., of the first and second node and for different $\varphi$ and $\theta$, as displayed in Figure \ref{fig:nodepos-polar} (a) and (b). Free fit parameters were $\aR$, $\bD$ and $q=g_{\perp} / g_{\parallel}$. The known parameters sheet carrier density $\ntd=8.51 \times 10^{11} \mathrm{cm^{-2}}$ determined from the periodicity of SdH-oscillations at $\theta=0^{\circ}$ and effective mass $m^*=0.37 m_0$ extracted from $T$-dependent measurements reported in \cite{Schaepers-1998} were fixed. The starting values for $g_{80}$ and $g_{xy}$ extracted from the coincidence measurements were taken in the first least squares fit. From the fit result, the corresponding coincidence angles $\theta_c^{1/2}$ at $\varphi=45^{\circ}$ and $\varphi=135^{\circ}$ at $\Bp=2.1 \, \mathrm{T}$ were calculated and compared to the results from coincidence measurements (see table \ref{tab:coincidence_angles}). This was found to overestimate the anisotropy of $\theta_c^{1/2}$ compared to experiment. This is expected and can be understood as follows: The actual value of $g_{xy}$ is smaller than the upper bound of $g_{xy}=0.14 $ when the residual influence of SOI terms on $\theta_c^{1/2}$ at $\Bp=2.1 \, \mathrm{T}$ is not negligible in the experiment. Therefore the fit routine was performed while systematically varying $g_{xy}$ between 0 and 0.14. The best fit was achieved for $g_{xy}=0.11$ and is shown as dashed lines in figure \ref{fig:nodepos-polar}. The extracted parameters together with their estimated uncertainties are:
\begin{equation}
  \label{eq:SOI-g-fitting-result}
\begin{aligned}
  \aR &=& (4.62 &\pm 0.09) \times 10^{-12} \, \mathrm{eVm}\\
  \bD &=& (0.46 &\pm 0.14) \times 10^{-12} \, \mathrm{eVm}\\
  g_{\perp} &=& -2.91 &\pm 0.60 \\
  g_{\parallel} &=& -4.72 &\pm 0.05 \\
  g_{xy} &=& 0.11 &\pm 0.02 \ .
\end{aligned}
\end{equation} Note that the relative sign of $\aR$ and $\bD$ is thus found to be positive as is the sign of $g_{xy}$.

At this point we would like to comment on one particular aspect of the above procedure: The in-plane anisotropy of $\tg$ on the one hand and the anisotropy due to the SOI terms on the other hand lead to beat node anisotropies with the same symmetry. So how can the fitting procedure distinguish the two components? Of course, independent measurement of $g_{xy}$ via the coincidence method at the highest available magnetic fields where the SOI-terms can be neglected is straightforward. However, we saw in the analysis above that in our case there \emph{was} a residual influence of the SOI terms in the coincidence measurements. The answer lies in the strongly different dependence of the Zeeman- and the SOI terms on the magnetic field. While $E_{\mathrm{Z}} \propto B$, it is $E_{\mathrm{R}}, E_{\mathrm{D}} \propto \sqrt{(n+1)B_{\perp}}$ which is roughly independent of B for high $n$. Our routine overcomes the problem by simultaneously fitting the $\varphi$-anisotropy of experimental data obtained in three different magnetic field regimes, namely the second node (at $B_{\perp} \approx 0.3$~T), the first node (at $B_{\perp} \approx 0.8$~T), and the coincidence (at $B_{\perp} \approx 2.1$~T). In these measurements in different field regimes the information is encoded that allows to separate the influence of $g_{xy}$ and the SOI, such that it is neither possible to model the data by assuming $g_{xy}=0$ and $|\bD|>0$, nor by assuming $|g_{xy}|>0$ and $\bD=0$. Both effects are clearly present in the experiment as is expected from theory \cite{Kalevich-1993}.

In the following we briefly comment on the contribution of the $k$-cubic term in the Dresselhaus part of the Hamiltonian that we did not include in the main part of this work. Following \cite{Das-1990}, $k$-cubic terms of the Dresselhaus Hamiltonian lead to a renormalization of $\bD$, such that we have to replace $\bD= -\gamma \langle k_z^2 \rangle$ in (\ref{eq:Hamiltonian}) by $\bD'=-\gamma(\langle k_z^2 \rangle - 1/4 k_{\text{F}}^2)$. Here $\gamma$ is the bulk Dresselhaus constant. Further, additional off-diagonal matrix elements $\langle n,+|H|n+3,-\rangle=-(i\gamma)/4 k_{\text{F}}^3$ appear \cite{Das-1990}, coupling states with quantum numbers $n$ and $n+3$. Using the parameters extracted from the experiment, we verified that the effect of these matrix elements on the results was negligible.

\section{Discussion}
\label{sec:discussion}

The result of the present experiment is the unambiguous extraction of $\aR$ and $\bD$ and their relative sign together with the full $\tg$-anisotropy. For this, no \emph{a priori} assumptions were necessary. The ability to determine all parameters at the same time provides a more comprehensive picture of spin-related phenomena in semiconductor heterostructures. In particular, $\mathbf{k} \cdot \mathbf{p}$-theory predicts $g_{xy} = \frac{2 \gamma e}{\hbar \mu_B} \left(\langle k_z^2 \rangle\langle z\rangle - \langle k_z^2 z \rangle \right) $. Our experimental approach determining $\bD'=-\gamma(\langle k_z^2 \rangle - 1/4 k_{\text{F}}^2)$ and $g_{xy}$ on an equal footing may thus pave the way to quantitatively validate the theoretical predictions. This is however, beyond the scope of the present paper. Our analysis highlighted the deep connection between $g_{xy}$ and the SOI-terms. We were able to disentangle the two contributions to the measured anisotropy in the numerical fitting routine, finding that the SOI terms had a significant influence on our coincidence measurements.  In this particular case at $B_{\perp} \approx 2.1$ T, $\approx 70 \%$ of the anisotropy in $\theta_c^{1/2}$ could be attributed to $g_{xy}$ while $\approx 30 \%$ were due to the influence of the SOI terms. Addressing the pure components of $\tg$ in a coincidence experiment would require higher magnetic fields. Our model predicts that for the specific system investigated here, the SOI contribution to $\theta_c^{1/2}$ becomes negligible at $B_{\perp} \approx 10$ T. At $\theta \approx 80^{\circ}$ this amounts to $B \approx 58$ T, which can be reached in pulsed magnetic fields at large scale facilities.

Our work reveals a remarkable out-of-plane anisotropy of $\tg$ with $g_{\perp}=-2.98$ and $g_{\parallel}=-4.72$. This result is consistent with previous experiments of Kowalski \emph{et al} on similar InGaAs heterostructures \cite{Kovalski-1994}. A significant out-of-plane anisotropy of $\tg$ was also predicted by $\mathbf{k} \cdot \mathbf{p}$-theory for InGaAs, albeit with $|g_ {\perp}|>|g_{\parallel}|$ \cite{Winkler-2003}. However, the correct way to derive $g_ {\perp}$ and $g_{||}$ is controversial \cite{Pfeffer-2006}. The small size of the in-plane $\tg$-anisotropy represented by $g_{xy}=0.11$ as compared to the diagonal components $g_{\perp}$ and $g_{\parallel}$ is similar to experiments on AlGaAs-based heterostructures \cite{Oestreich-1996,Eldridge-2011}. The outcome of a dominant Rashba parameter $\aR = (4.62 \pm 0.09) \times 10^{-12} \, \mathrm{eVm}$ is consistent with previous experiments on the same heterostructures \cite{Schaepers-1998,Rupprecht-2013}. Furthermore a smaller $\bD = (0.46 \pm 0.14) \times 10^{-12} \, \mathrm{eVm}$ on the order of $0.1 \aR$ is consistent with the more recent analysis in Ref.~\cite{Herzog-2015}, where the oscillatory $\mu(\Bp)$ was measured in weakly tilted fields.

\section{Conclusion}
\label{sec:conclusion}

In conclusion, we have determined the Rashba- and Dresselhaus constants $\aR = (4.62 \pm 0.09) \times 10^{-12} \, \mathrm{eVm} $ and $\bD = (0.46 \pm 0.14) \times 10^{-12} \, \mathrm{eVm}$ and their relative sign (positive) together with the full $g$-tensor $\tg$ with $g_{\perp}=(-2.91 \pm 0.60)$, $g_{\parallel}=(-4.72 \pm 0.05)$ and $g_{xy}=(0.11 \pm 0.02)$ for an asymmetric InGaAs quantum well. For this we employed SdH magnetotransport measurements in doubly tilted magnetic fields together with numerical calculations of the SOI-induced beat node positions. The method presented here is especially powerful to explore systems where one of the two components is small: If one of the two SOI terms is exactly zero, the node positions are isotropic. The breaking of this symmetry is a very sensitive gauge capable of resolving the smaller term.

\ack
\label{sec:acknowledgements}
We thank E I Rashba and D Maslov for valuable discussions and gratefully acknowledge financial support by the DFG TRR80 and via SPP1285 ``semiconductor spintronics'', Grant No. GR1640/3 as well as the German Excellence Initiative via Nanosystems Initiative Munich (NIM).

\section*{References}

\bibliographystyle{iopart-num}
\bibliography{literature}
\providecommand{\newblock}{}
\end{document}